\documentclass[twocolumn,journal]{IEEEtran}

\usepackage{graphics} 
\usepackage{xcolor}
\usepackage{multirow}
\usepackage{epsfig}
\usepackage{graphicx}
\usepackage{siunitx}
\usepackage[T1]{fontenc}
\usepackage[latin9]{inputenc}
\usepackage{textcomp}
\usepackage{graphicx}
\usepackage{xcolor}
\usepackage[unicode=true,
 bookmarks=true,bookmarksnumbered=true,bookmarksopen=true,bookmarksopenlevel=1,
 breaklinks=false,pdfborder={0 0 0},pdfborderstyle={},backref=false,colorlinks=false]
 {hyperref}
\hypersetup{pdftitle={SuperGaN: Synthesis of NbTiN/GaN/NbTiN Tunnel Junctions},
 pdfauthor={Your Name},
 pdfpagelayout=OneColumn, pdfnewwindow=true, pdfstartview=XYZ, plainpages=false}

\makeatletter

 \let\oldforeign@language\foreign@language
 \DeclareRobustCommand{\foreign@language}[1]{%
   \lowercase{\oldforeign@language{#1}}}

\usepackage[caption=false,font=footnotesize]{subfig}

\begin{document}
\title{SuperGaN: Synthesis of NbTiN/GaN/NbTiN Tunnel Junctions}

\author{Michael Cyberey,~\IEEEmembership{Senior Member,~IEEE,}, Scott Hinton, Christopher Moore, Robert M. Weikle,~\IEEEmembership{Fellow,~IEEE,} and Arthur W. Lichtenberger,~\IEEEmembership{Senior Member,~IEEE} 
\thanks{Michael Cyberey, Scott Hinton, Christopher Moore, Robert M. Weikle, and Arthur W. Lichtenberger are with the Department of Electrical Engineering, the University of Virginia, Charlottesville, VA, USA.
        { mcyberey@virginia.edu}
        This work has been submitted to the IEEE for possible publication. Copyright may be transferred without notice, after which this version may no longer be accessible.
        }}

\markboth{1-MP-TF-01S - 16$^{th}$ European Conference on Applied Superconductivity}
\IEEEpubid{}
\maketitle
\begin{abstract}

Nb-based circuits have broad applications in quantum-limited photon detectors, low-noise parametric amplifiers, superconducting digital logic circuits, and low-loss circuits for quantum computing. The current state-of-the-art approach for superconductor-insulator-superconductor (SIS) junction material is the Gurvitch trilayer process based on magnetron sputtering of Nb electrodes with Al-Oxide or AlN tunnel barriers grown on an Al overlayer \cite{gurvitch_high_1983}. However, a current limitation of elemental Nb-based circuits is the low-loss operation of THz circuits operating above the ~670 GHz gap frequency of Nb and operation at higher temperatures for projects with a strict power budget, such as space-based applications.

NbTiN is an alternative higher energy gap material and we have previously reported on the first NbTiN/AlN/NbTiN superconducting-insulating-superconducting (SIS) junctions with an epitaxially grown AlN tunnel barrier \cite{cyberey_nbtin/aln/nbtin_2019}.  One drawback of a directly grown tunnel barrier compared to thermal oxidation or plasma nitridation is control of the barrier thickness and uniformity across a substrate, leading to variations in current density (J$_c$).  Semiconductor barriers with smaller barrier heights enable thicker tunnel barriers for a given J$_c$ \cite{kroger_niobium_1981}. GaN is an alternative semiconductor material with a closed-packed Wurtzite crystal structure similar to AlN and it can be epitaxially grown as a tunnel barrier using the Reactive Bias Target Ion Beam Deposition (RBTIBD) technique.  This work presents the preliminary results of the first reported high-quality NbTiN/GaN/NbTiN heterojunctions with underdamped SIS I(V) characteristics.

\end{abstract}


\IEEEpeerreviewmaketitle{}

\section{Introduction}

\IEEEPARstart{N}{}b-based circuits have broad applications ranging from quantum sensors to classical and quantum computing applications. A notable limitation of Nb-based circuits is the 2.8 meV energy gap of Nb, limiting low-loss operation to roughly 670 GHz and temperatures of 4 K or lower. A high energy gap superconducting material such as NbN or NbTiN is required for higher-frequency applications or operations at higher temperatures. We have previously reported on our development of NbTiN films and NbTiN/AlN/NbTiN superconducting-insulating-superconducting (SIS) junctions using the novel RBTIBD technique with superconducting energy gaps as high as 5.5 meV \cite{farrahi_development_2019}. The RBTIBD technique combines a low-energy few-eV plasma that avoids unwanted sputtering with independent pulse biasing up to 2 kV of the target materials \cite{zhurin_biased_2000}. This results in precise control of the ion impact energy and resulting adatom energetics, allowing the tuning of deposition parameters to realize smooth films with surface roughness comparable to molecular beam epitaxy (MBE) and reduced intermixing at a superconducting-insulating interface \cite{cyberey_growth_2018}. The low surface roughness, combined with precise control of energies at the film's surface, allows the realization of high-quality heterojunctions.

 One limitation of physical vapor deposition (PVD) grown tunnel barriers is the control of thickness and uniformity across a substrate, leading to variations in J$_c$. Semiconductor barriers with smaller barrier heights than AlO$_x$ or AlN enable thicker tunnel barriers for a given J$_c$. The J$_c$ control and uniformity of thicker tunnel barriers is impacted less by variations in barrier thickness, and thicker tunnel barriers are less likely to contain appreciable pinholes and defects. Wurtzite phase GaN is a semiconductor with a ~2\% lattice mismatch to FCC NbTiN in the close-packed plane; it is a potential alternative barrier material to AlN to realize NbTiN-based SIS junctions with a lower bandgap semiconductor barrier. This work presents the preliminary results of the first high-quality NbTiN/GaN/NbTiN heterojunctions with underdamped SIS I(V) characteristics.

\begin{figure}
\begin{centering}
\includegraphics[width=1\columnwidth]{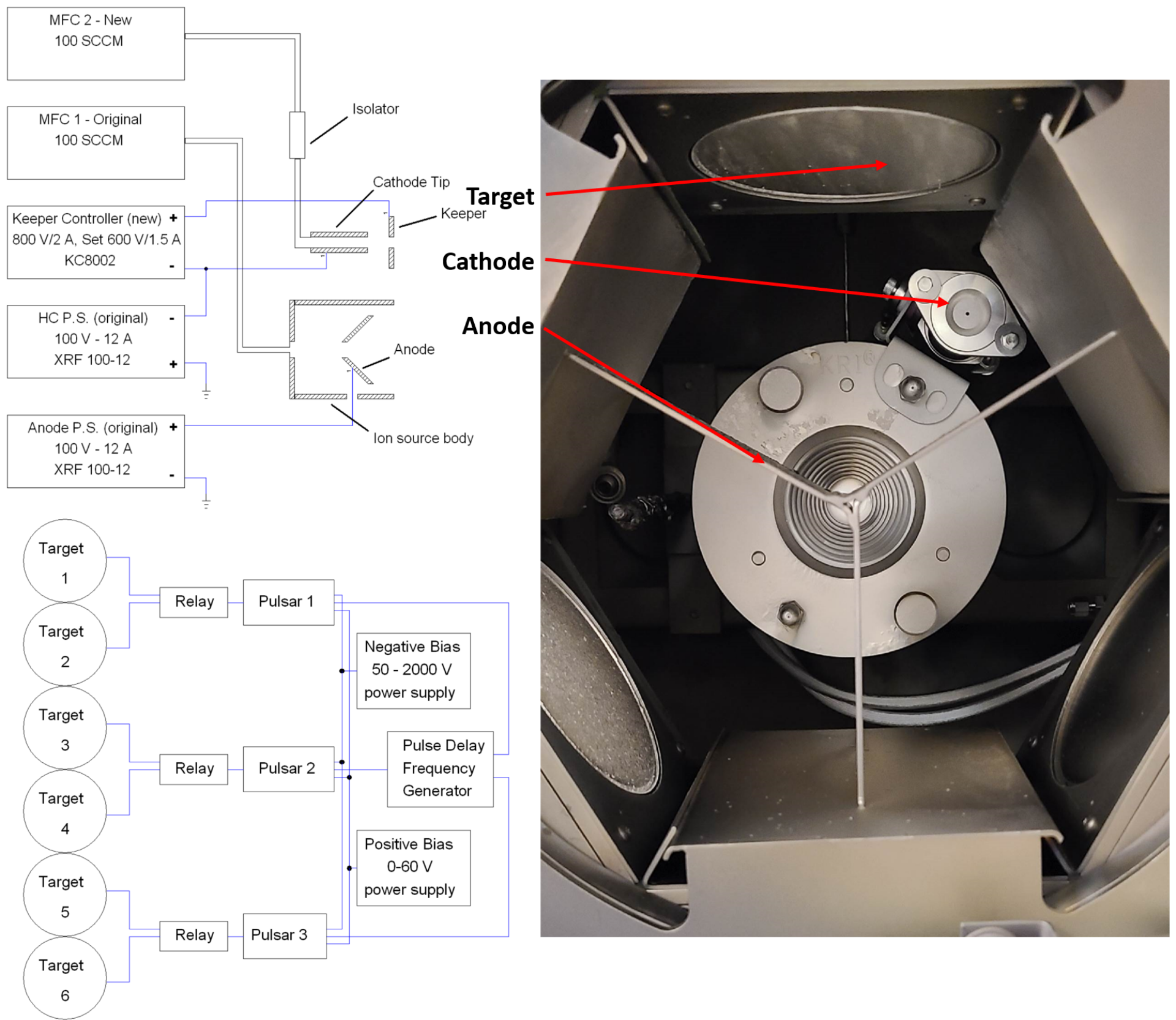}
\par\end{centering}
\caption{\label{fig:LANS}Schematic and photograph of our modified RBTIBD system.}
\end{figure}

\begin{figure*}
    \begin{centering}
        \includegraphics[width=.8\columnwidth]{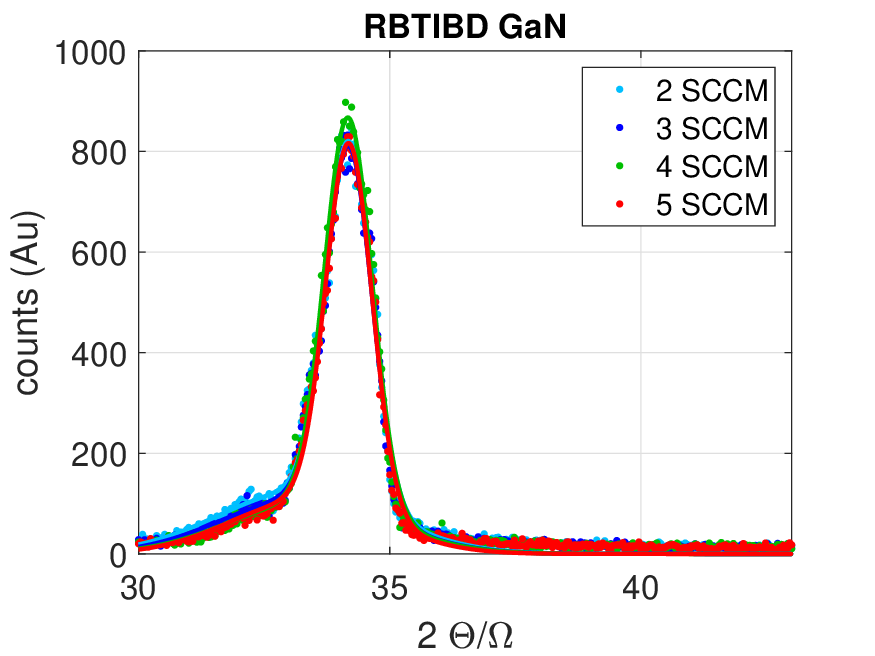}
        \includegraphics[width=1\columnwidth]{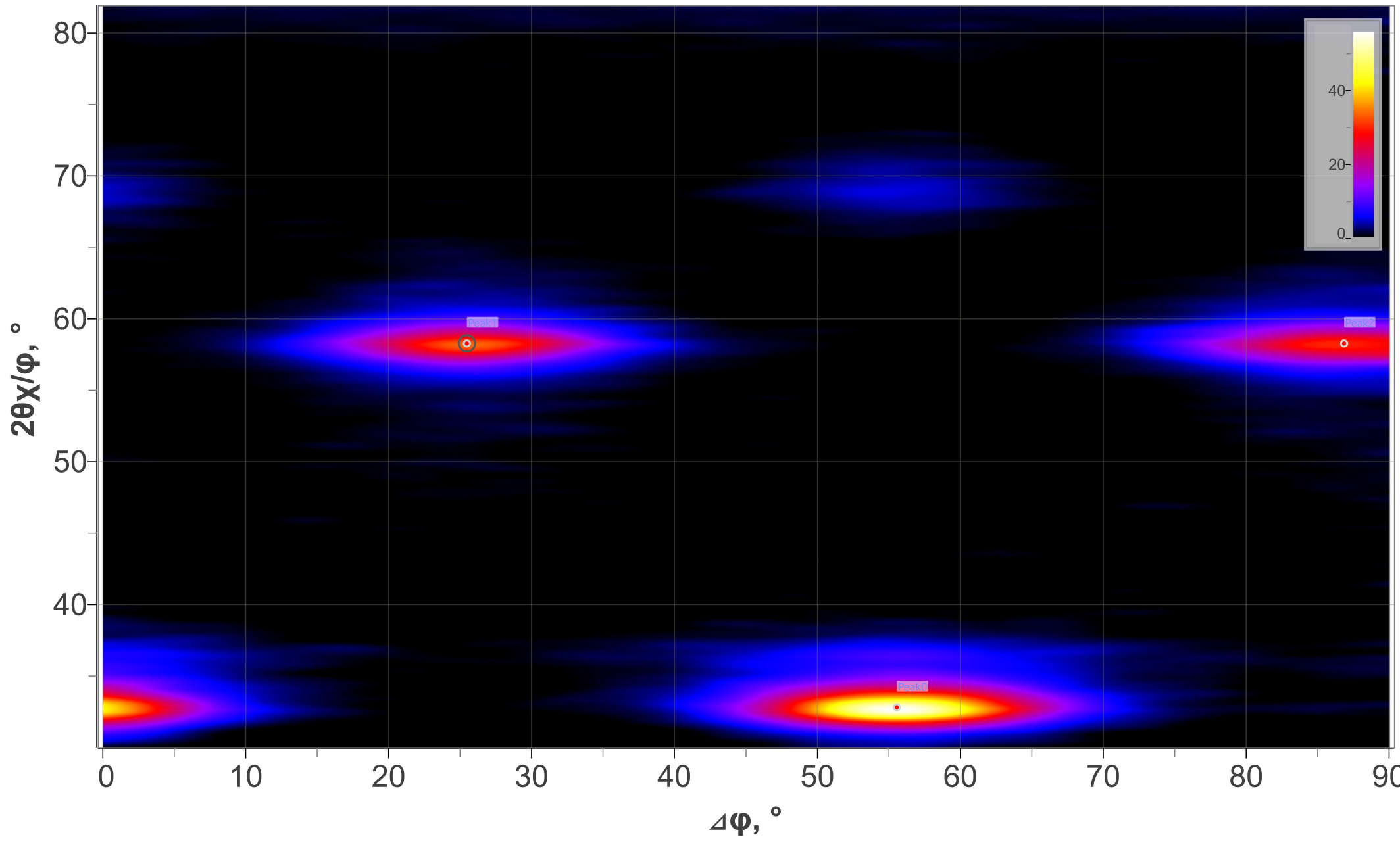}
    \par\end{centering}
        
    \caption{\label{fig:XRD}: Left: 2$\theta$-$\omega$ scan showing (002) peak of GaN films at varying N$_2$ flow rates, Right: Reciprocal space mapping (RSM) with an in-plane XRD geometry showing peaks from the (100) and (110) planes}
\end{figure*}

\begin{figure}
    \begin{centering}
        \includegraphics[width=1\columnwidth]{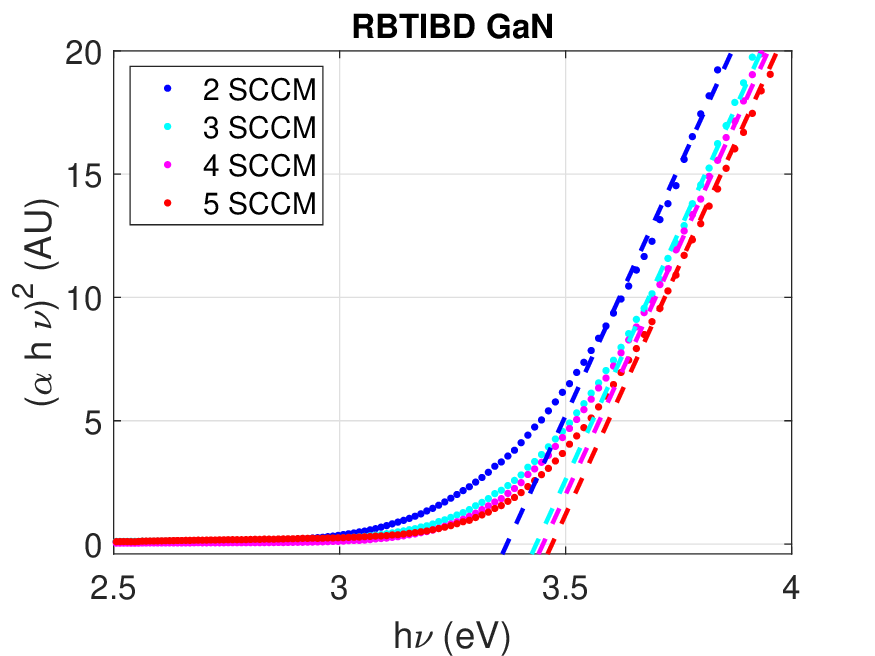}
        \par\end{centering}
    \caption{\label{fig:Egap} Shown is ${(\alpha h\nu)}^{(1/n)}$ as a function of $h\nu$ for RBTIBD GaN films. The E$_{g}$ is determined from the x-intercept of the linear region and increases monotonically with N$_2$ flow rate from left to right on the graph.}
\end{figure}

\begin{figure*}
    \begin{centering}
        \includegraphics[width=1\columnwidth]{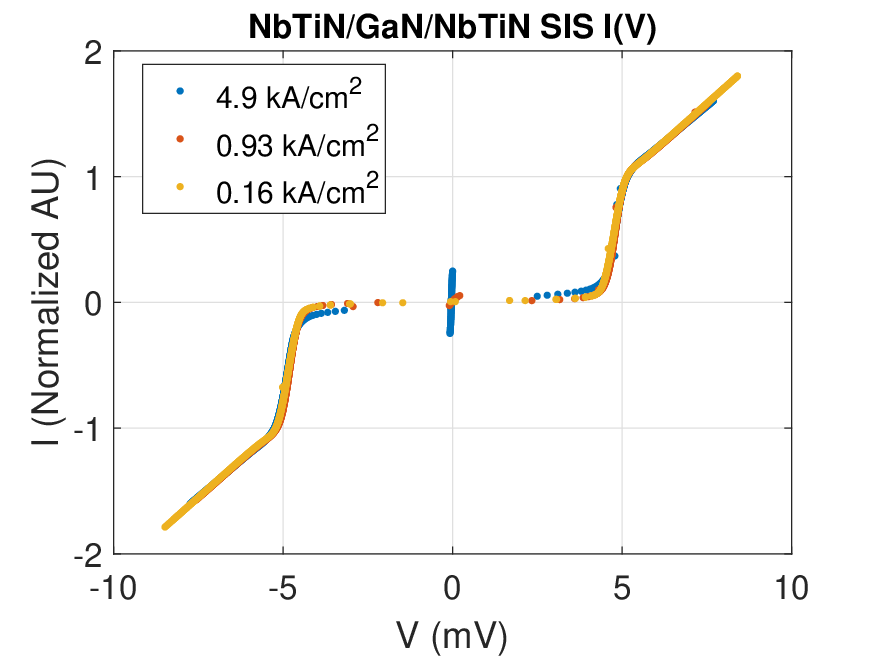}
        \includegraphics[width=1\columnwidth]{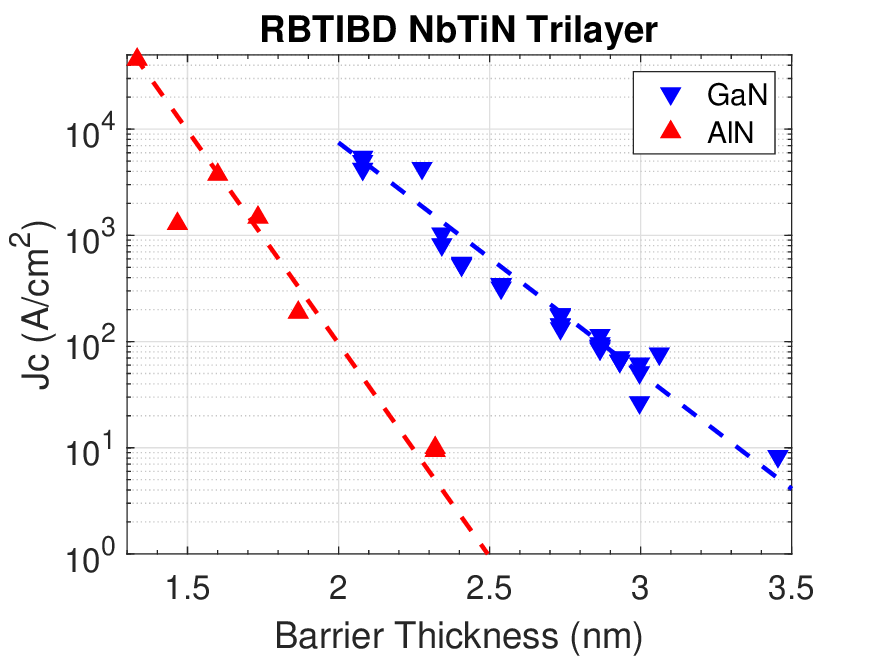}
    \par\end{centering}
        
    \caption{\label{fig:SIS}: Left: Typical I(V) characteristics of our NbTiN/GaN/NbTiN junctions, Right: J$_c$ as a function of barrier thickness for both GaN and AlN tunnel barriers}
\end{figure*}

\section{Experimental Setup}
All films were grown using our modified RBTIBD system, shown in Fig. \ref{fig:LANS}. Our RBTIBD tool is a versatile platform for multilayer material synthesis, epitaxial growth, and compositional nanotechnology. It offers a hybrid approach between ion beam and conventional sputter deposition. It utilizes a hollow-cathode ion source to generate a low-energy (5-50 eV) broad beam plasma and independent control of the incoming flux and dynamics to each target through pulsed biasing. The system contains six sputtering targets, three of which can be simultaneously co-sputtered, with control of the deposition rates adjusted through the duty cycle and frequency of the bias pulses. The system can notably realize films and heterojunctions with surface roughness and interface quality comparable to MBE.

Bulk GaN films of 80-160 nm nominal thickness were grown by RBTIBD of a 100 mm diameter 99.99\% purity GaN target at room temperature on C-plane sapphire substrates. A 7 A ion current with 10 SCCM of Ar gas flow was used for the hollow cathode, and a 6.5 A load was applied to the anode with an additional 70 SCCM of Ar gas flow. A target bias of -900 V was applied as a 71 KHz pulse bias with 3 $\mu$s duty cycle, and +20 V was applied to the target during the off cycle to prevent charge accumulation on the semiconducting target. N$_2$ gas was introduced to the anode before film growth and varied from 2 to 5 SCCM. Crystallographic analysis of our GaN films for 2$\theta$-$\omega$, 2$\theta$-$\phi$, and Reciprocol Space Mapping (RSM) was performed using a Rigaku Smartlab XRD. Optical properties were measured using a JA Woolaam M-2000D ellipsometer with a wavelength ranging from 193 to 1690 nm. Film thickness measurements were performed using the standard stylus profilometer technique.

NbTiN/GaN/NbTiN trilayer films were grown on thermally oxidized (100) Si substrates using previously established RBTIBD NbTiN growth techniques and the RBTIBD GaN conditions established for bulk films with 5 SCCM N$_2$ gas flow. The GaN tunnel barrier thickness was varied from 2 to 3.5 nm. SIS junctions were fabricated with a diameter ranging from 0.6 to 3.0 $\mu$m using our self-aligned rapid SIS junction fabrication technique described elsewhere \cite{cyberey_investigation_2014}. Four-point I(V) measurements were performed at 4 K in a custom-built closed cycle Gifford-McMahon (GM) based cryostat.

\section{Results and Discussions}
\subsection{GaN Film Properties}

The $2\theta$-$\omega$ XRD spectra for GaN films deposited at varying N$_2$ gas flow are shown in Fig. \ref{fig:XRD}. Compared to NbTiN growth from a NbTi target, the XRD spectra peak position and FWHM in our $2\theta$-$\omega$ scans did not appreciably change with varying N$_2$ flow rate, and only 2-3\% N$_2$ gas flow was needed to realize Wurtzite-phase GaN, likely due to the use of a compound GaN target.  We performed RSM by performing in-plane 2$\theta$-$\phi$ scans while varying $\phi$ from 0 to 90 degrees; the resulting RSM is shown in Fig. \ref{fig:XRD}. Dots are observed for diffraction of the (100) and (110) planes with expected 6-fold symmetry with the lack of Debye rings; this indicates local or partial epitaxial growth of (001) oriented GaN films with slight in-plane rotational misalignment.
The energy gaps of our GaN films were calculated using the optical absorption coefficients measured using our M2000 ellipsometer in transmission mode by a method proposed by Tauc through the relationship $\alpha h\nu = \alpha_0(h\nu-E_g)^n$, where $\alpha$ is the absorption coefficient, $\alpha_0$ is a constant, $h\nu$ is the photon energy, $E_g$ is the energy gap, and the exponent n denotes the nature of the transition; a value of 0.5 for n denotes a direct transition while a value of 1 denotes an indirect transition \cite{tauc_optical_1968}. Through plotting ${(\alpha h\nu)}^{(1/n)}$ as a function of $h\nu$, one can determine the energy gap from the x-intercept of the fit to the linear region of this plot as shown in Fig. \ref{fig:Egap} for our RBTIBD GaN films. The direct energy gap of our films varies from 3.37 to 3.47 eV with a monotonic increase to the $N_2/Ar$ flow ratio and is in good agreement with other reports in the literature for GaN films grown by PVD \cite{ueno_optical_2019,thao_reactively_2020}.

\subsection{NbTiN/GaN/NbTiN Junction Properies}

Fig. \ref{fig:SIS} shows typical I(V) characteristics of the fabricated SIS tunnel junctions for the realized J$_c$ range. The subgap resistance to normal resistance ratio ($R_{sg}/R_n$) was greater than 10 for all measured junctions. The current density as a function of barrier thickness is plotted on the right of in Fig. \ref{fig:SIS}. The expected $J_c=Ae^{Bt}$ relationship is observed, and compared to our previously reported NbTiN/AlN/NbTIN tunnel junctions, a thicker GaN barrier layer is required to realize a given J$_c$ value. This is due to the lower barrier height of GaN compared to AlN.

\section{Conclusions}
In this work, we continued our investigation and the characterization of NbTiN-based tunnel junctions using room-temperature RBTIBD by exploring an alternative semiconductor tunnel barrier. We have demonstrated room-temperature GaN films with (100) orientation with local epitaxial growth. We realized NbTiN/GaN/NbTiN SIS tunnel junctions with directly deposited GaN tunnel barriers with sum-gap voltages approaching 5.0 mV and an exponential relationship between J$_c$ and barrier thickness. This junction technology opens future research for devices with higher operational frequency and higher operational temperature than Nb-based devices. We will further investigate GaN growth at elevated temperatures and alternative semiconductor materials. We also note that we recently replaced our original ion source with an improved custom solution, allowing twice as much ion current and should further improve film quality.

In summary, we have demonstrated as a proof-of-concept and reported, for the first time, high-quality RBTIDB synthesized NbTiN/GaN/NbTiN SIS junctions with energy gaps approaching 5.0 meV.  The $R_{sg}/R_n$ ratio remained greater than 10 even for our highest J$_c$ of 5000 A/cm$^2$.  Future work will further decrease the GaN tunnel barrier thickness to explore higher J$_c$ values

\section*{Acknowledgments}

This work was supported by the National Radio Astronomy Observatory, and National Ground Intelligence Center. The National Radio Astronomy Observatory is a facility of the National Science Foundation operated under cooperative agreement by Associated Universities, Inc. 

\bibliographystyle{IEEEtran}
\bibliography{References}

\begin{IEEEbiographynophoto}{Michael Cyberey}
(Senior Member, IEEE) received his Ph.D. degree in electrical engineering from the University of Virginia, Charlottesville, VA, USA, in 2014. He is currently a Principal Scientist with the Department of Electrical and Computer Engineering, University of Virginia, and his research interests include novel microfabrication techniques, THz devices, spintronics, nanomaterials, and superconducting electronics including superconducting tunnel junction devices, kinetic inductance detectors, and traveling wave kinetic inductance parametric amplifiers. Dr. Cyberey is a fabrication engineer for Dominion MicroProbes, Inc and founder of NanoMEC, LLC.\end{IEEEbiographynophoto}

\begin{IEEEbiographynophoto}{Scott Hinton}
Scott Hinton is a Ph.D. candidate in electrical engineering at the University of Virginia. His research focus is on microwave design, periodic structures, and the modeling, fabrication, and measurement of low-noise superconducting amplifiers.  He is a student member of IEEE.\end{IEEEbiographynophoto}

\begin{IEEEbiographynophoto}{Christopher Moore}
Christopher Moore is a Ph.D. candidate in electrical engineering at the University of Virginia in the Charles L. Brown Department of Electrical and Computer Engineering. His research focus is on submillimeter electronics, non-linear circuit design, microfabrication, vacuum technology, and devices. He is a student member of IEEE, an amateur radio operator, and an electronic warfare expert. \end{IEEEbiographynophoto}

\begin{IEEEbiographynophoto}{Robert M. Weikle, II}
 Robert M. Weikle, II received his Ph.D. from Caltech in 1992 and is a Professor in the Charles L. Brown Department of Electrical and Computer Engineering at the University of Virginia, where he serves as Director of the Far-Infrared and Terahertz Receiver Laboratory. Professor Weikle's research focuses on millimeter and submillimeter-wave electronics, sensors, device fabrication, characterization, and circuit design. Dr. Weikle  has been PI or Co-PI on approximately \$40 million in funded research programs and an author on over 240 technical papers in refereed journals and proceedings. He is a  Fellow of the IEEE and a founding member of Dominion MicroProbes, Inc.
\end{IEEEbiographynophoto}

\begin{IEEEbiographynophoto}{Arthur W. Lichtenberger}
Arthur W. Lichtenberger,  received his  Ph.D. in electrical engineering from the University of Virginia in 1987 and is a Professor in the Charles L. Brown Department of Electrical and Computer Engineering at the University of Virginia, where he serves as director of the UVA  Innovations in Fabrication (IFAB) cost-share user facility. His research expertise encompasses superconducting materials, devices, and circuits in conjunction with submillimeter electronics, high-frequency instrumentation and metrology. To date, he has been PI or Co-PI on over 60 million dollars of funding and an author on over 230 papers. Dr. Lichtenberger is a founding member of Dominion Microprobes, Inc.\end{IEEEbiographynophoto}

\end{document}